# Deep Feed-Forward Neural Network for Bangla Isolated Speech Recognition

**Dipayan Bhadra[1], Mehrab Hosain[2], and Fatema Alam[3]**

**Abstract:** As the most important human-machine interfacing tool, an insignificant amount of work has been carried out on Bangla Speech Recognition compared to the English language. Motivated by this, in this work, the performance of speaker-independent isolated speech recognition systems has been implemented and analyzed using a dataset that is created containing both isolated Bangla and English spoken words. An approach using the Mel Frequency Cepstral Coefficient (MFCC) and Deep Feed-Forward Fully Connected Neural Network (DFFNN) of 7 layers as a classifier is proposed in this work to recognize isolated spoken words. This work shows 93.42% recognition accuracy which is better compared to most of the works done previously on Bangla speech recognition considering the number of classes and dataset size.

## 1. Introduction

Speech recognition is the most important interfacing tool for human-machine interaction. A lot of works have been done on the English speech recognition system and it has been reached to the state of the art. On the other hand, implementation of speech recognition system for Bangla language is a growing field of research. Approximately 235 million native speakers and another 40 million as second language speakers speak in Bangla language making it the sixth most spoken native language and the seventh most spoken language by the total number of speakers in the world. It is the official language of the states of West Bengal and Tripura in India. Since it is the seventh most spoken language, Bangla speech recognition system helps a lot of people. Research on Bangla came into focus in the 90's. Recognition of Bangla speech has been started since around 2000 (Sultana, et. al. 2021). In 2002 A. Karim et al. presented a method for Spoken Letters Recognition in Bangla (Karim et. al., 2002). In the same year K. Roy et al. presented the Bangla speech recognition system using Artificial neural networks (Roy et.

[1] Power Grid Company of Bangladesh Limited, Email: dipu_dipayan@outlook.com
[2] Louisiana Tech University, Email: robinhosain@gmail.com
[3] Jahangirnagar University, Email: fatemazinnah89@gmail.com

al., 2002). In 2003, M.R. Hassan presented a phoneme recognition system using Artificial Neural Network (ANN) (Hasan et.al., 2003) and in the same year K.J. Rahman presented a continuous speech recognition system using ANN in 2003 (Rahman et. al., 2003). Later beside ANN, various algorithms like Dynamic Time Wrapping (DTW), Hidden Markov Model (HMM), etc., have been developed over time as classifiers for Bangla speech recognition system (Noman & Cheng, 2022; Sultana, et. al. 2021). Moderate progress has been made in Bangla Speech Recognition since 2007. In 2007, Hasnat et. al. proposed MFCC coefficients and HMM and stochastic language model and got an accuracy rate of 85% for 100 words recognition system (Hasnat, et. al., 2007). In 2009, Ghulam Muhammad et al. developed an HMM-based speaker-independent Bangla digits recognizer which used their own dataset of 10000 words recorded from 50 males and 50 females (Muhammad & Alotaibi, 2009). In 2013, Ali et.al. used Gaussian Mixer Model and MFCC as features and got 84% accuracy on 1000 classes problem (Ali et. al., 2013). In the same year, Hossain et. al. proposed MFCC and ANN with 1 hidden layer-based speech recognition system for 10 classes problem and got 92% accuracy. In 2016, Ahammad and Rahman used MFCC and ANN with 1 hidden layer and got 98.46% accuracy on 10 classes problem (Ahammad & Rahman, 2016). In 2022, Noman & Cheng used ANN with 1 hidden layer as a classifier and absolute values of DFT as features for 7 classes problem and got an accuracy of 95.23% (Noman & Cheng, 2022). In the same year, in the work of Sen et. al. MFCC spectrogram and CNN of 6 layers were used for a 100 classes problem and the accuracy rate was 89.61%

Inspired by the effectiveness of MFCCs as features, in this work, a speech recognition system is proposed for speaker-independent Bangla Isolated speech recognition system using 7 layers deep feed-forward neural network for 60 classes problem. A dataset has been created with the help of 25 different speakers from different zones of Bangladesh. 36 Bangla and 24 English words have been captured to make a dataset size of 1800 words to evaluate the performance of the proposed system. In this work, the depth of the fully connected neural network is more than ever reported for Bangla speech recognition and the result of this work is showing better recognition accuracy compared to the most of the works done previously on Bangla speech recognition considering the number of classes, dataset size etc

**2. Materials and Methods**

**2.1 Dataset creation**

In this work, a dataset is created containing 36 Bangla words and 24 English words with the help of 25 different persons from different regions of Bangladesh. 30 samples per word from both male and female speakers are recorded. The number of speech samples in the created dataset is 1800. The samples are recorded using a sound recorder (smartphones) in room environment. Among them 1200 samples are used for the training and 600 samples are used for testing. All the samples are recorded as ".wav" file. The speech sample words are recorded with a high sampling frequency (44.10 kHz) to create this dataset. In time domain, one example is shown in Figure 1.

### 2.1.1 Words used to create the dataset: The words used are [Bangla word (pronunciation, meaning)]:

দাঁড়াও (Stand), হাঁটো (Walk), থামো (Stop), সামনে (Forward), পিছনে (Backward), ডানে (Right), বামে (Left), উপরে (Up), নিচে (Down), শুরু (Start), শেষ (End), পড় (Read), লিখ (Write), শুনো (Listen), বলো (Speak), তাকাও (Look), আলো (Light), শব্দ (Sound), সময় (Time), ভর (Mass), নাম (Name), বই (Book), খাতা (Pad), কলম (Pen), গাড়ি (Car), বাড়ি (House), পানি (Water), খাবার (Food), কুকুর (Dog), বিড়াল (Cat), মানুষ (Human), শিশু (Child), হাতুড়ি (Hammer), চশমা (Glasses), গান (Song), ঘড়ি (Watch), Light, Mobile, Internet, WiFi, Class, Program, Diode, Capacitor, Switch, Television, Radio, Head-phone, Mike, Battery, 1, 2, 3, 4, 5, 6, 7, 8, 9, 0.

**Table 1:** Database size and distribution

| Training Sample Size | Test Sample Size | Total Dataset Size |
|---|---|---|
| 1200 Sample | 600 Sample | 1800 Sample |
| 66.67% | 33.33% | 100% |

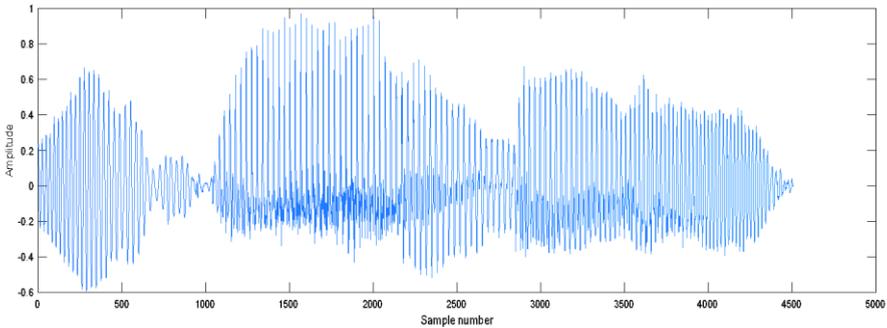

**Figure 1:** Speech signal of "বলো" under the dataset in time domain. There are 29,520 samples values (points) in the word. Here we can see the silence portion of the speech (signal) that at the beginning of the signal as well as the end.

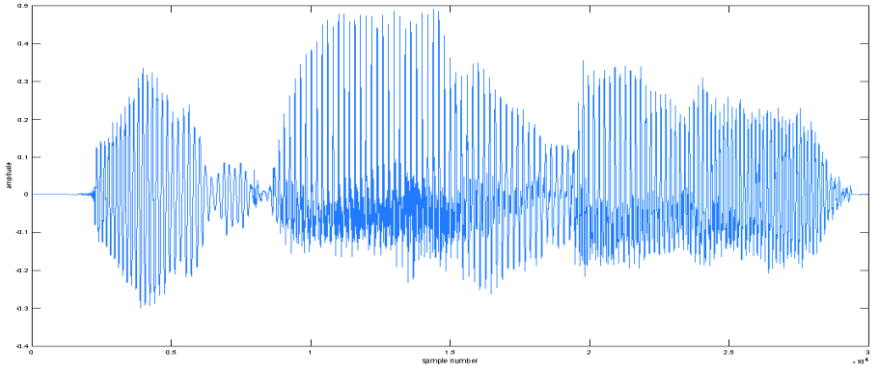

**Figure 2:** The normalized, down-sampled signal "বলো" after silence removal. There are only 4,510 sample values (points) in the word.

## 2.2 Preprocessing

Because of the high sampling frequency, the data points are very high in a speech sample (signal). Down sampling is required to reduce the data size and computational cost. Also, due to the difference in volume of speech of different words from different speakers as well as different words of same speaker, normalization is necessary. Hence, pre-processing of the recorder samples is done before taking features from the signals. Down sampling, normalization, silence removal by voice activity detection, windowing and pre-emphasis are done in the preprocessing stage of speech signal in this work.

### 2.2.1 Down Sampling

The speech sample words are recorded with a high sampling frequency (44.10 kHz) to create this dataset. Considering important harmonic frequencies, human speech is generally below 5 kHz. So the recorded signals (words) with 44.1kHz are down-sampled at 10 kHz. Every sample point in a signal is stored as 8 bytes, so there are 256 levels of a sample point amplitude.

### 2.2.2 Normalization

Normalization is done by dividing the sample values of every speech signal with the maximum value present in that speech signal.

### 2.2.3 Voice Activity Detection and Silence Removal

The silence portion of a signal is useless for recognition, because it contains no information. The voiced part of a speech signal is detected based on the energy level of that signal within a window of 108 millisecond. And then a threshold is used to determine whether it is a voiced portion or silence portion. After that, the silence portion is removed. After that, the silence portion is deleted using matrix manipulation. Figure 2 shows the normalized and down-sampled signal "বলো" after silence removal.

### 2.2.4 Windowing

In an audio signal, the statistical properties over time are not constant. But they are assumed to be constant for a very short interval of time. The time for which the signal is considered for processing is called a window, and the data acquired in a window is called a frame. In this work, Hamming window with the frame length of 300 and the 100 sample frame shift is used. Hamming window is used because it introduces the least distortion. 3/4 frame overlap is considered. The sampling frequency is 10000 Hz. So one frame is 300/10000 = 30 ms and the frame shift is 100/10000 = 10 ms.

### 2.2.5 Pre-emphasis

Pre-emphasis is done by a first-order high-pass filtering, difference equation in time domain is: Emphasis Signal (n) = Signal(n) − a ∗ Signal(n − 1)

Or as a transfer function in z-domain: H (z) = $(l - a \cdot z^{-1})/l$; in this work '**a**' is set to 0.9375.

### 2.3 Feature Extraction

After pre-processing, compressed MFCCs are used as features and fed as the input to a deep feed-forward neural network. Here k-means clustering is used as feature compression technique.

### 2.3.1 MFCC

MFCC is always considered to be the best available approximation of human ear. MFCC is a representation of the short-term power spectrum of a sound, based on a linear cosine transform of a log power spectrum on a nonlinearmel scale of frequency. In this work 40 filters are used. With 10000Hz sampling frequency, the number of filter banks is 40 in this work along with frame length: 300 and the frame overlap is 200. The result is a N×14 matrix of MFCCs for each N frame. The number N will naturally vary for each voice sample.

### 2.4 Feature Compression

k-means classifier is used to cluster the vectors that contain the MFCCs to get compressed feature vectors. In this work, used 'K' value is 5 & 8 to analyze the performance. The final value of K is 8. The result is a 8×14 matrix, instead of the previous N by 14 matrix. We further reshape the matrix into a 1 by 112 (=8×14) vector, which is the input to the neural network.

**2.5 DFFNN for classification:**

Fully connected feed-forward neural network is used as a classifier for this work. Sigmoid function is used as activation function of the neurons and backpropagation is used as training algorithm for the NN. The final output is achieved from a 7 layers (1 input, 5 hidden and 1 output) neural network. Number of neurons in input layer is 112 (=8×14) and number of neurons at the output layer is 60. But a lot of networks with different depth have been analyzed. In this work, different numbers of hidden layer are used. 1 to 6 hidden layers are tested for performance analysis. Also different numbers of nodes or neurons in hidden layer are tested: 25 to 100.

Network 1: total 3 layers (one hidden layer), different number of neurons used for hidden layers: 25, 35, 45, 55, 65, 70

Network 2: total 4 layers (two hidden layers), different number of neurons used for hidden layers: (45, 35), (45, 45), (45, 55), (50, 45), (55,45)

Network 3: total 5 layers (three hidden layers), different number of neurons used for hidden layers: (40, 40, 40), (50, 55, 50)

Network 4: total 6 layers (four hidden layers), different number of neurons used for hidden layers: (45,25,25,45), (45, 45, 45,45), (60, 70, 80, 30), (75, 45, 80, 50), (45, 35, 35, 45), (50, 40, 80, 100), (65, 55, 45, 35), (80, 50, 70, 60), (90, 90, 90, 90), (100, 100, 100, 100)

Network 5: total 7 layers (five hidden layers), different number of neurons used for hidden layers: (85, 85, 90, 85, 85), (75, 85, 90, 85, 85), (85, 85, 90, 85, 95), (75, 85, 90, 105, 115), (90, 90, 90, 90, 90), (85, 85, 86, 85, 85), (85, 85, 90, 85, 95), (65, 65, 70, 70, 90), (85, 90, 85, 90, 90), (85, 85, 85, 85, 85), final network: (100, 95, 90, 95, 100),

Network 6: total 8 layers (six hidden layers), different number of neurons used for hidden layers: (85, 85, 90, 85, 85, 85)

Figure 3 represents the block diagram of the proposed scheme showing the series of steps under preprocessing, feature extraction, feature compression and classification. In this task, the speech samples have been collected in low-noise environment, so no denoising steps has been implemented in this task to clean the signal from environmental noise.

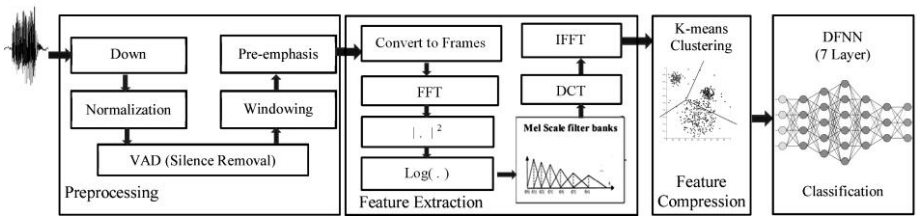

**Figure 3:** Block Diagram of the proposed scheme

## 1. RESULTS AND DISCUSSION

### 3.1 Experimental platform

#### 3.1.1 Hardwire Configuration

Processing Unit: Intel Core i7 7700K Processor (7th Gen.), Ram: 8GB DDR4, GPU: NVIDIA GTX 960 with 4 GB dedicated RAM

#### 3.1.2 Software Configuration

Operating System: Windows 10 Professional, 64 bit Computational Software: MATLAB 2018a 64 bit

### 3.2 Parameters of the proposed system

★ MFCC as features (compressed into 1×112 vector size with K-means classifier with K=8)

- Number of Cepstral Coefficients: 14
- Number of filter banks: 40
- Frame length: 300 samples
- Frame overlap: 200 samples

★ Weight and Bias initialization method: Gaussian distribution random numbers

★ Learning rate considered: 0.05, Decay factor used: 0.95, Batch size: 1 sample

★ Activation function used: Sigmoid, Cost function used: Squire Error (Euclidean distance)

★ Neural Network architecture is: (112, 100, 95, 90, 95, 100, 60)

★ Input layer: 112 neurons, Output layer: 60 neurons, hidden layers =5

★ Total parameters (Neural network weights and biases): 53,840

★ 60 class problems, test accuracy 93.42%

### 3.2 Accuracy on test dataset

**Table: 2:** Accuracy on test dataset

| Sample Information | Test set Accuracy (%) |
|---|---|
| No of Class=60, Test Sample =600 | 93.42% |

## 3.4 Comparative Analysis of the performance of proposed method with some preivious prominent works

**Table 4:** Comparison with some reported work of Bangla speech recognition:

| Sl. No. | Work Reference | Database information | Feature Extraction & Classifier | Reported Accuracy |
|---|---|---|---|---|
| 1. | (Ahammad & Rahman, 2016) | * Bangla Digits (10 classes)<br>* No info about dataset size. | * MFCC Feature Extraction<br>* ANN with one hidden layer | **98.46%** |
| 2. | (Ali, et al., 2013) | * 1000 words, 10 samples for each word Total: 10000 samples | * MFCC<br>* Gaussian Mixer Model<br>* Posterior Probability Function | **84%** |
| 3. | (Hossain, et al. 2013) | *10 classes, 10 speakers, 150 training samples, 150 test samples | *MFCC features<br>*ANN with 1 hidden layer | **92%** |
| 4. | (Muhammad & Alotaibi, 2009) | * Bangla Digits (10 classes)<br>* 370 samples as training set<br>* 130 samples as test set.<br>Total: 5000 samples. | * MFCC<br>* Gaussian Mixer Model<br>* HMM classifier | **93.6%** |
| 5. | (Hasnat, et al., 2007) | * 100 words<br>* No info about dataset size. | * MFCC<br>* HMM classifier | **85%** |
| 6. | (Noman & Cheng, 2022) | *7 classes (words)<br>* 28 training samples<br>* no other info about dataset | * absolute value of DFT<br>*ANN with one hidden layer | **95.23%** |
| 7. | (Sen, et al., 2022) | *100 classes, 100 samples for each class=10000 words<br>*40,000 samples created | *MFCC spectrogram<br>* CNN (4 Conv layer, 1 dense, 1 output layer) | **89.61%** |
| 8. | (Paul, et al., 2022) | * 10 digits, 4 words, 100 samples/word<br>Total: 1400 samples. | *MFCC as features, DTW for classification | **93%** |
| 9. | Proposed Method | * 60 words, 30 samples/word<br>Total: 1800 samples.<br>(1200 samples for training and 600 samples for testing) | * MFCC<br>*DFFNN (7 Layers) | **93.42%** |

In the work of Ahammad & Rahman, 2016, the reported accuracy is 98.46%. But the number of classes is only 10 which is far less than this work. Also, in the work of Muhammad & Alotaibi, 2009, no of classes is 10 and the reported accuracy is 93.6%, which is almost equal to the accuracy of the proposed method. In the work of Hossain, et al., in 2013 only 10 class problem was considered and still, the accuracy rate is less than that of the proposed method. Noman & Cheng in 2022 got an accuracy of 95.23% but the work was only 7 class problem with very small dataset size. In 2022, Paul et al. studied 14 class problems and got 93% accuracy rate which is slightly less than the proposed method. In the other works (Sen, et. al., 2022; Hasnat et al., 2007 and Ali, et al., 2013) the no of classes is more than our work, but the recognition accuracy rate is considerably lower than that of the proposed method. We can see, that the proposed system outperforms other prominent works on Bangla speech recognition considering database size, number of classes, and recognition accuracy.

### 3.5. Comparative Analysis of the performance of DFFNN with different neuron numbers and compressed MFCCs as input features

**Table 3:** Analysis of the performance of DFNN with different number of neurons

| Hidden Layer and Features | | No. of Neurons | Epoch | Learning Rate, ETA | Training Accuracy |
|---|---|---|---|---|---|
| 5 hidden layer (MFCCs as feature) | K=5 | 65,65,70,70,90 | 50K | 0.05 | 87% |
| | K=5 | 85,90,85,90,90 | 50k | 0.05 | 88.41% |
| | K=8 | 85,90,85,90,90 | 50k | 0.05 | 91.08% |
| | K=8 | 85,85,85,85,85 | 60k | 0.06 | 91.75% |
| | **K=8** | **100, 95, 90, 95, 100** | **100k** | **0.05** | **93.75%** |

Among all the network architectures, 7 layer neural network provided the best training accuracy in all the trials.

### 4. CONCLUSION

In this work, not only a moderate dataset has been created but also this work proposes a recognition model that gives better recognition accuracy compared to the most of the works done previously on Bangla speech recognition. Among all the variations of Deep Neural Networks along with LPCC and MFCC as features, the network with 5 hidden layers seems to provide the best output with MFCC as features compressed by K-means classifier with K=8 in all the trials. We can see, that proposed system outperforms other prominent works in terms of database size, number of classes, and recognition accuracy. There is an insignificant difference between the training accuracy (93.75%) and the test accuracy (93.42%) indicating that the neural network classifier is not overfitted during the training. The words selected during the development of the dataset are pronounced very differently from each other. Also, the proposed system can identify words pronounced by both male and female speakers equally well. This indicates the robustness of the proposed system in real-world applications. The main drawback of the proposed system is the absence of denoising filters because the signals have been recorded in low-noise environments. The proposed system may not perform equally well if the input is a very noisy signal with unwanted speech signals as a background- further investigation is required in this direction.

There are still some limitations that need to be investigated in the future. The prospects for future work are:

Signals that are used in this work are recorded in low-noise environments, so they can be regarded as pure signals. In practice, noise will make a difference in the accuracy of detection. Noise removal algorithms or networks can be utilized in recognition of the noisy speech signals.

Other recent developments of neural network architectures like variations of Deep Convolutional Neural Networks can be tested for further improvements in speech recognition tasks. Since CNN does not require a feature extraction step, it can be applicable to real-time systems involving speech, speaker, or emotion recognition. Recent state-of-the-art recognition systems are using deep CNN and some systems are outperforming humans in terms of recognition accuracy.

**References**

Ahammad, K., & Rahman, Md. M. 2016. Connected Bangla Speech Recognition using Artificial Neural Network. *International Journal of Computer Applications,* 149(9), 38–41. DOI: 10.5120/ijca2016911568

Ali, M. A., Hossain, M., & Bhuiyan, M. N. 2013. Automatic Speech Recognition Technique for Bangla Words. *International Journal of Advanced Science and Technology*, 50, 51–59. http://article.nadiapub.com/IJAST/vol50/5.pdf

Hasan, M.R., Hasan, M.M., Hossain, M.Z. 2021. How many Mel to frequency cepstral coefficients to be utilized in speech recognition? A study with the Bengali language. The J. of Eng. 817–827. https://doi.org/10.1049/tje2.12082

Hassan M.R., Nath B., Bhuiyan M.A. 2003. Bengali phoneme recognition: a new approach. In: Proc. 6th Int. conf. on computer and information technology (ICCIT03).

Hasnat, M. A., Mowla, J. & Khan, M. 2007. Isolated and Continous Bangla Speech Recognition: Implementation. *Performance and Application Perspective Center for Research on Bangla Language Processing*. http://dspace.bracu.ac.bd:8080/xmlui/.....

Hossain, M. A., Rahman, M. M., Prodhan, U. K. & Khan, M. F. 2013. Implementation Of Back-Propagation Neural Network For Isolated Bangla Speech Recognition. *IJIST*-3(4), 1–9. DOI: 10.5121/ijist.2013.3401.

Muhammad, G. & Alotaibi, Y. 2009. Automatic Speech Recognition for Bangla Digits. *Proceedings of 2009, 12th International Conference on Computer and Information Technology (*ICCIT 2009, 379–383. DOI: 10.1109/ICCIT.2009.5407267

Karim A.R., Rahman MS, Iqbal MZ. 2002. Recognition of Spoken Letters in Bangla. In: In proceedings of the 5th ICCIT conference. ICCIT. p. 1–5

Noman, A. & Cheng, X., 2022. Bengali Isolated Speech Recognition Using Artificial Neural Network. *Mechatronics and Automation Technology, 14 to 23*. DOI: 10.3233/ATDE 221144

Paul, B., Paul, R., Bera, S., & Phadikar, S. (2022). Isolated Bangla Spoken Digit and Word Recognition Using MFCC and DTW. *In Engineering Mathematics and Computing* (pp. 235-246). Singapore: Springer Nature Singapore. https://doi.org/10.1007/978-981-19-2300-5_16

Rahman K.J., Hossain M.A., Das D., Islam A.Z.M.T., Ali D.M.G. 2003. Continuous Bangla Speech Recognition System. In: Proc. 6th Int. Conf. on Computer and Information Technology (ICCIT03). p. 1–5.

Roy K., Das D., Ali M.G. 2002. Development of the speech recognition system using artificial neural network. Proc. 5th Int. conf. on computer and information technology (ICCIT02). p. 118–122.

Sen, O., Roy, P. and Mahmud, A. 2022. A Novel Bangla Spoken Numerals Recognition System Using Convolutional Neural Network. *SSRN* http://dx.doi.org/10.2139/ssrn.4127691

Sultana, S., Rahman, M. S., & Iqbal, M. Z. 2021. Recent advancement in speech recognition for Bangla: A survey. Int. J. Adv. Comput. Sci. Appl, 12(3), 546–552 https://pdfs.semanticscholar.org/9267/...